 \documentclass[final,5p,times,twocolumn,authoryear]{elsarticle}

\usepackage{amssymb}
\usepackage{lipsum}

\journal{New Astronomy}

\usepackage{tikz}
\usetikzlibrary{calc}
\usetikzlibrary{decorations.pathmorphing}
\usepackage{etoolbox}

\usepackage{hyperref}
\hypersetup{
    colorlinks=true,
    citecolor=blue,
    linkcolor=blue,
    urlcolor=blue
}

\makeatletter
\patchcmd{\@cite}{\@citea\hskip\@citeh\@hyper@linkstart{cite}{\@citeb}\@citeb\@hyper@linkend}{\@citea\textcolor{blue}{\@citeb}}{}{}
\makeatother

\usepackage{sectsty}
\usepackage{natbib}
\usepackage{amsmath}
\usepackage{array}
\usepackage{adjustbox}
\usepackage{amssymb}
\usepackage{multicol}
\usepackage{float}
\usepackage{multirow}
\usepackage{gensymb}
\usepackage[nopatch]{microtype}
\usepackage{booktabs}

\begin{document}

\begin{frontmatter}



\title{Modelling the scattering by porous aggregate dust grains in the Far-Ultraviolet halos of Spica and Achernar}


\author[first]{Nilanjana Dey Choudhury}
\author[first]{P. Shalima} \ead{shalima.p@manipal.edu}
\author[first]{Keerthana U.}
\author[second]{J. Murthy}
\affiliation[first]{organization={Manipal Centre for Natural Sciences},
            addressline={Manipal Academy of Higher Education},
            state={Karnataka},
            city={Manipal},                       
            postcode={576 104},
            country={India}}

\affiliation[second]{organization={Indian Institute of Astrophysics},
            addressline={Koramangala}, 
            city={Bangalore},
            postcode={560034}, 
            state={Karnataka},
            country={India}}           
\begin{abstract}
Far-Ultraviolet (FUV) halos have been detected around six bright stars by \citet{Murthy2011_UV_halos} using GALEX observations. These halos are thought to be caused by forward scattering of the starlight by dust grains present in thin foreground clouds. The optical constants of grains producing such halos have been constrained earlier by using a single scattering model, that considered the Henyey-Greenstein empirical phase function instead of theoretical phase functions for the scattering grains.
In this work, we have modelled the FUV halos for two stars, Spica and Achernar, by considering the realistic porous aggregates of different sizes and compositions. 
As the Henyey-Greenstein phase function is known to deviate from theoretical predictions, 
we have utilized theoretical scattering phase functions for modelling.
The dust is placed in a double-layered plane-parallel sheet with its distance and optical depth varied to get the best fit.
We find that the halo intensities are dominated by scattering due to 0.05 $\mu m$ sized porous dust aggregates made of amorphous silicate and carbonaceous aggregates for Spica and Achernar, respectively. 
We find that the medium in front of Achernar has a lower optical depth ($\tau$) of 0.032 compared to Spica which has a value of $\tau$ = 0.1. This low value is close to the optical depth of the local ISM (0.01) within 40 pc of the Sun. This study demonstrates an effective method to constrain the dust grain properties in the local interstellar medium.
\end{abstract}




\end{frontmatter}




\section{Introduction}
\label{introduction}

\citet{Murthy2011_UV_halos} detected far-ultraviolet (FUV) halos around six bright stars with intensities ranging from 200 to 5000 photon units (1 photon unit = 1 ph/cm$^{2}$/s/sr/\AA) from {\it{GALEX}} observations. The forward scattering of star light by the thin foreground dust clouds is considered to be responsible for these halos. They are ideal candidates for exploring the FUV scattering phase function of interstellar dust grains at smaller angles. \citet{Murthy2011_UV_halos} modeled the halo observations for 4 bright stars: Spica, Achernar, Algenib and Mirzam. The remaining two stars, Sirius and Adhara were not modeled since they did not have sufficient foreground material.  
They were able to uniquely constrain the phase function asymmetry parameter ($g$) of the grains, but due to uncertainty in dust grain geometry they were unable to constrain the albedo ($a$). 

Since the dust is distributed into multiple components along the line of sight,  \citet{Murthy2011_UV_halos} modeled the halos using a double sheet geometry which consists of two plane-parallel dust clouds placed at two different distances from the star.
\citet{Shalima2013}  separately modeled the Northern and Southern regions in UV halo around Spica, by considering the fact that the Spica halo is not symmetric. They  obtained the best-fit values of $a$ and $g$ to be 0.26 $\pm$ 0.1 and 0.58 $\pm$ 0.11 for the Northern regions of Spica and constrained the optical depth ($\tau$) and dust cloud distance ($d$) to 0.047 $\pm$ 0.006 and 3.65 $\pm$ 1.05 pc respectively. Assuming the dust grain optical properties to be the same for both regions, they constrained the $\tau$ and $d$ for Southern regions to be 0.04 $\pm$ 0.01 and 9.5 $\pm$ 1.5 pc respectively. 
Previously, \citet{Murthy2011_UV_halos} and \citet{Shalima2013} considered the Henyey-Greenstein (HG) phase function (\citet{Henyey_Greenstein_1941}) for the dust scattering model. The HG phase function is a simple empirical function with a single parameter $g$. 

However, it deviates considerably from the theoretical phase function in the UV and IR (\citet{Baes2022}).
In this work, we aim to constrain the scattering characteristics, especially the scattering phase function of the dust grains that are responsible for the FUV halos of Spica and Achernar by using realistic phase functions computed for porous aggregates. 

\citet{Draine2003_Scattering}, \citet{draine2021} and \citet{Hensley2023} provided comprehensive dust grain models and corresponding phase functions for different size distributions and compositions of spherical grains that could reproduce the observed variations in the extinction curves.   
Apart from the theoretical phase functions, \citet{Hensley2023} has also provided an empirical phase function dependent on two parameters which is an improvement over the HG function.

However, from polarization measurements it has been shown that dust grains in the ISM are not necessarily spherical but elongated and irregularly shaped. They could be porous aggregates made up of smaller particles which are bound together to form voids or spaces between them. The presence of voids in porous aggregates increases the surface area of dust particles which leads to the enhancement in the scattering efficiency.
\citet{Mathis1989}, \citet{Dorschner1995} had modeled the porous aggregates as a cluster of small spheres called monomers and their optical properties were calculated by using different methods such as DDA and GMM. Both these methods have been used by \citet{Blum1998} and \citet{Blum2000} to derive the phase functions of grains. \citet{2008YueShen} considered porous aggregates for the study of light scattering properties of grains and their impact on interstellar extinction.

As part of this work, we consider a single scattering model in which the scattering grains are porous aggregates and validate the model against the observed FUV halo intensities of Spica and Achernar to constrain the grain properties. Since graphite and silicates are the most common constituents of dust grains in the ISM, we consider the aggregates to be composed of these two types. 
Finally, we also compare our results with those obtained using the HG phase function and the theoretical phase functions of \citet{Draine2003_Scattering} for spherical grains made up of a mixture of graphite and silicate. 

\section{Data}
\begin{figure}[H]
    \centering
    \includegraphics[width=8cm, height=7cm]{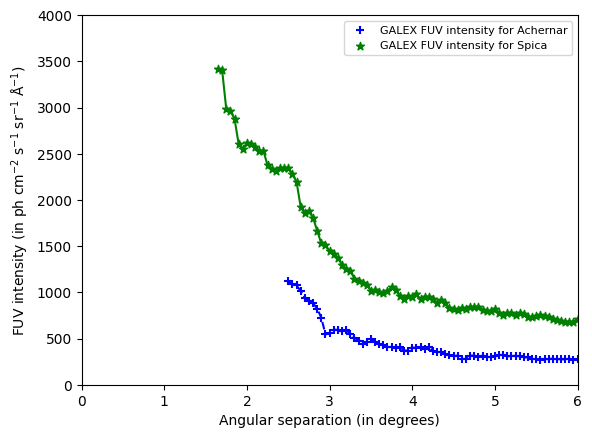}
    \caption{Variation of the observed GALEX FUV intensity for Spica and Achernar w.r.t angular separation}
    \label{fig:observation}
\end{figure}

We have used the All-Sky data from Galaxy Evolution Explorer ($GALEX$) spacecraft obtained by \citet{Murthy2011_UV_halos} for modeling the scattering intensities in FUV. As $GALEX$ couldn't observe too close to the bright stars, the nearest observation data available is about 2° away from Spica and Achernar (see fig.\ref{fig:observation}). The details about the data analysis is mentioned in \citeauthor{Murthy2011_UV_halos} (\citeyear{Murthy2011_UV_halos}).

\section{Model}

We have modeled the FUV halo intensities around Spica and Achernar by considering single scattering due to dust grains that are porous aggregates.  
Two different compositions are used, i.e. Silicates and Carbonaceous grains, to derive the best-fit model. Along with typical compositions of ISM, graphite and astronomical silicates as given by \citet{1984BTDraine}, amorphous carbon and amorphous silicate \citep{2022Demyk} are also considered in the model . 
The details of our model are outlined in the following sections.

\subsection{Scattering model}
We have modeled the FUV intensities in the stellar halos around two bright stars: Spica and Achernar by utilizing the observations  from \citet{Murthy2011_UV_halos}.
The main advantage of these two locations is that they have maximum number of observations and the radiation field in the halo is dominated by the central bright star, making them easier to model.
We consider single scattering due to the low column density values observed for Spica and Achernar. The single scattering intensity ($I_{sca} $) is calculated (in photon units) according to the following equation:
\begin{equation}
    I_{sca} = \frac{L}{4 \pi r^2 } \times { a} \times {\phi(\theta)}\times  {\tau}
    \label{eq:I-sca}
\end{equation}

Here, $a$ is the albedo, \textit{L} is the luminosity of the star, $r$ is the distance (in pc) between the star and the dust cloud, $\phi(\theta)$ is the scattering phase function which is dependent on the scattering angle denoted by $\theta$ and $\tau$ is the optical depth of the scattering layer.
Rather than using the Henyey-Greenstein phase function we have used theoretical phase functions corresponding to the porous aggregate grain model, described in the next section. 

\subsection{Grain model}
\begin{figure}[H]
\centering
\includegraphics[height=5.0cm,width=5.0cm]{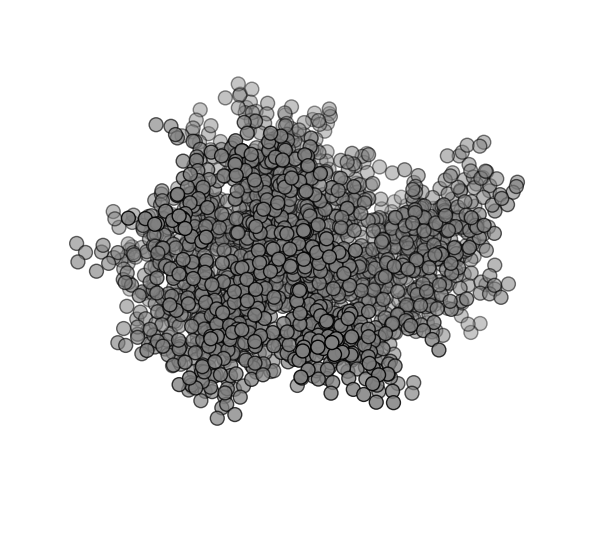}
    \caption{Aggregate containing 2048 monomers}
    \label{aggr}
\end{figure}

In the current study, we use DDSCAT package to determine the phase function and cross-sections for the aggregates having four different compositions as listed in Table \ref{prop}. The properties of aggregates are calculated by keeping the dipole sites in an ambient medium, i.e. vacuum. The grain model contains 9490 dipoles out of 50 $\times$ 50 $\times$ 75 dipole sites. As seen in the Table \ref{prop}, three different grain sizes are considered for computing the scattering properties at a particular wavelength $\lambda$ = 1521 \AA. Here, the porosity is fixed at 86.44\%. The geometry of the aggregate is taken from \cite{2008YueShen} and the structure containing 2048 monomers is shown in Fig. \ref{aggr}.

\begin{table}[H]
\caption{Properties of Aggregates}
\vspace{0.2cm}
\resizebox{0.5\textwidth}{!}{%
\begin{tabular}{|c|c|c|c|c|c|c|c|}
\hline
\textbf{Aggregate} &\textbf{No. of} & \textbf{$\lambda$} & \textbf{Aggregate} & \textbf{ Monomer} & \textbf{Albedo} & \textbf{Asymmetry} \\
\textbf{type}&\textbf{monomers}& &\textbf{size} &\textbf{ size} & &\textbf{parameter}\\
&&(\AA)&($\mu$m) & \textbf{(nm)}& \textbf{($a$)}& \textbf{($g$)}\\

\hline
\multirow{3}{*}{}{Astronomical Silicate} & \multirow{3}{*}{}{2048}
&\multirow{3}{*}{}{1521}  & 0.01 & 0.79 & 0.09 & 0.08  \\
\cline{4-7}
& & & 0.05  & 3.94 & 0.53 &  0.85  \\
\cline{4-7}
& & & 0.1 & 7.87 & 0.65 &  0.92  \\

\hline
\multirow{3}{*}{}{ Graphite (a)}& \multirow{3}{*}{}{2048} &\multirow{3}{*}{}{1521}  &  0.01 &  0.79  & 0.11 &  0.08  \\
\cline{4-7}
& & & 0.05 & 3.94  & 0.60 &  0.85  \\
\cline{4-7}
& & & 0.1 & 7.87 & 0.73 &  0.92  \\

\hline
\multirow{3}{*}{}{ Graphite (b)}& \multirow{3}{*}{}{2048} &\multirow{3}{*}{}{1521}  &  0.01 & 0.79  & 0.0026 & 0.0079    \\
\cline{4-7}
& & & 0.05 & 3.94  & 0.25 & 0.86   \\
\cline{4-7}
& & & 0.1 & 7.87 & 0.41 & 0.93   \\

\hline
\multirow{3}{*}{}{Amorphous Carbon} & \multirow{3}{*}{}{2048} &\multirow{3}{*}{}{1521}  &  0.01  & 0.79 & 0.056 & 0.079    \\
\cline{4-7}
& & & 0.05 & 3.94 & 0.416 & 0.857   \\
\cline{4-7}
& & & 0.1 & 7.87 & 0.557 & 0.926   \\

\hline
\multirow{3}{*}{}{Amorphous Silicate}& \multirow{3}{*}{}{2048} &\multirow{3}{*}{}{1521}  & 0.01 & 0.79 & 0.077 & 0.079    \\
\cline{4-7}
\multirow{3}{*}{}{(Forsterite)} & & & 0.05 & 3.94  & 0.492 & 0.855   \\
\cline{4-7}
& & & 0.1 & 7.87 & 0.612 & 0.917   \\

\hline
\multirow{3}{*}{}{Amorphous Silicate}& \multirow{3}{*}{}{2048} &\multirow{3}{*}{}{1521}  & 0.01 & 0.79 & 0.121 & 0.079    \\
\cline{4-7}
\multirow{3}{*}{}{(Enstatite)}& & & 0.05 & 3.94  & 0.613 & 0.850   \\
\cline{4-7}
& & & 0.1 & 7.87 & 0.703 & 0.902   \\
\hline
\end{tabular}
}
\label{prop}
\end{table}
\subsection{Dust Distribution}
\begin{figure}[H]
    \includegraphics[height=7cm,width=7cm]
    {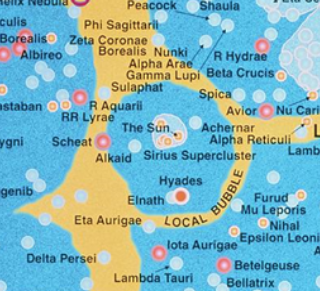}
    \caption{ Representation of the position of Spica and Achernar w.r.t interface between Local bubble and Loop I Super shell , \textit{Credit}: \citet{Zaninetti2020_local_bubble_shape}. }
    \label{fig:loop-bubble}
\end{figure}

\subsubsection{Spica}
Spica (l = 316.11°, b = 50.84°) also named as Alpha Virginis is a binary star of spectral type B1 III-IV which is situated inside the loop I Super shell as being mentioned in fig \ref{fig:loop-bubble}. It is located at a distance of 80 pc from Earth and surrounded by an HII region. The hydrogen column density in front of Spica is very low ($\sim$ $10^{19}$ cm$^{-2}$) but still results in the bright UV halo. Although the hydrogen column density measurements were available for Spica, we do not use it for the halo modeling as the observations were made 2° ($\sim$ 2.8 pc) away from Spica. 

We utilized the stellar flux for Spica convolved with the instrumental calibration, computed using the Kurucz model spectra  (\citet{Kurucz1979_model_atmospheres}) by \cite{Murthy2011_UV_halos} as the input to our model.
\citet{Park2010_HII_region} showed that the dust geometry is complicated and asymmetric around the Southern regions of Spica due to the presence of the interaction zone between loop I Super shell and local bubble. This interaction zone consists of shocks, resulting in the shattering of the larger grains. From the studies done by \citet{Centurion1991} and \citet{Zagury1998_Dust_composition}, it was deduced that dust grains found in medium towards Spica i.e. the interface between the local bubble and loop I Supershell are likely to be smaller in size.  However, according to \citet{Choi2013}, Spica could be beyond the Local bubble boundary, but not the interaction zone due to its low column density.

\begin{figure}[H]
    \includegraphics[width=7.5cm]{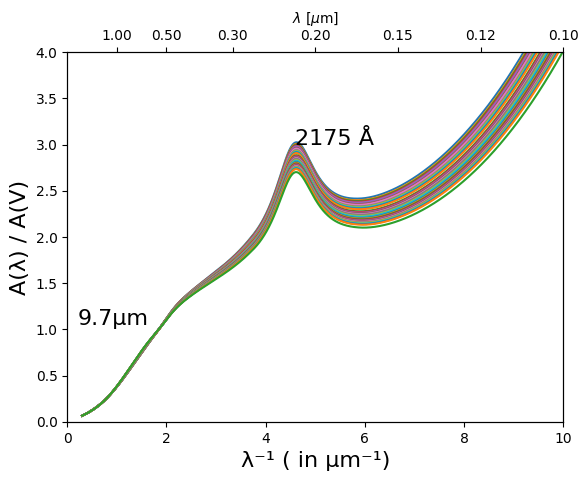}
    \caption{Extinction curves obtained for Stars mentioned in table \ref{tab:stars-near-Spica} near Spica using $R_V$ dependent Extinction law by \citet{Fitzpatrick1999}}
    \label{fig:F99-ext-curves-Spica}
\end{figure}

We follow \citet{Murthy2011_UV_halos} and consider the dust to be in the form of two layers in front of Spica. \citet{Murthy2011_UV_halos} derived the scattering layer distance to be 3 pc from Spica using 60/100 micron ratio.  Instead of keeping the scattering layer fixed at 3 pc, we vary the distance of the first scattering layer ($dl_{1}$), from 1 to 10 pc to obtain a best-fit distance for each location. The 2nd sheet is placed at 40 pc ($dl_{2}$), and it provides an offset to the intensity by contributing 20 to 30\%  of the scattered intensity.
From the studies of extinction measurements done by \citet{Jones2011_White_Dwarf_Spectra} and \citet{Puspitarini2012}, it was found that sufficient material exists within 90 pc from the star, with an optical depth ($\tau$) of 0.1. Therefore, we have distributed the material among the two scattering layers such that 20\% of the material is in the 1st sheet and remaining 80\% in the 2nd sheet. \\
In order to compare our results with the appropriate grain models of \citet{Draine2003_Scattering}, we require the extinction curves for Spica or other nearby stars within 200 pc.
Stars which are within 80 pc are considered to be present inside the Loop I super shell and near the scattering layer of Spica, therefore the dust grain properties and composition could be similar.
The stars within 200 pc, but beyond 90 pc are located beyond the Loop I Super shell and their extinction properties could represent the material in the interaction zone i.e. medium between Loop I Super shell and Local Bubble.
We find 48 stars in the region and their extinction parameters are given in Appendix 1, Table \ref{tab:stars-near-Spica}. The extinction curves for these stars are plotted in fig. \ref{fig:F99-ext-curves-Spica} using the $R_V$ dependent extinction law of \citet{Fitzpatrick1999}. We see that the R$_{V}$ parameter is close to 3.1, hence we consider Draine's model for R$_{V}$=3.1 to compare our models.
\begin{figure*}[t]
    \centering
    \includegraphics[height=8.0cm,width=16cm]
    {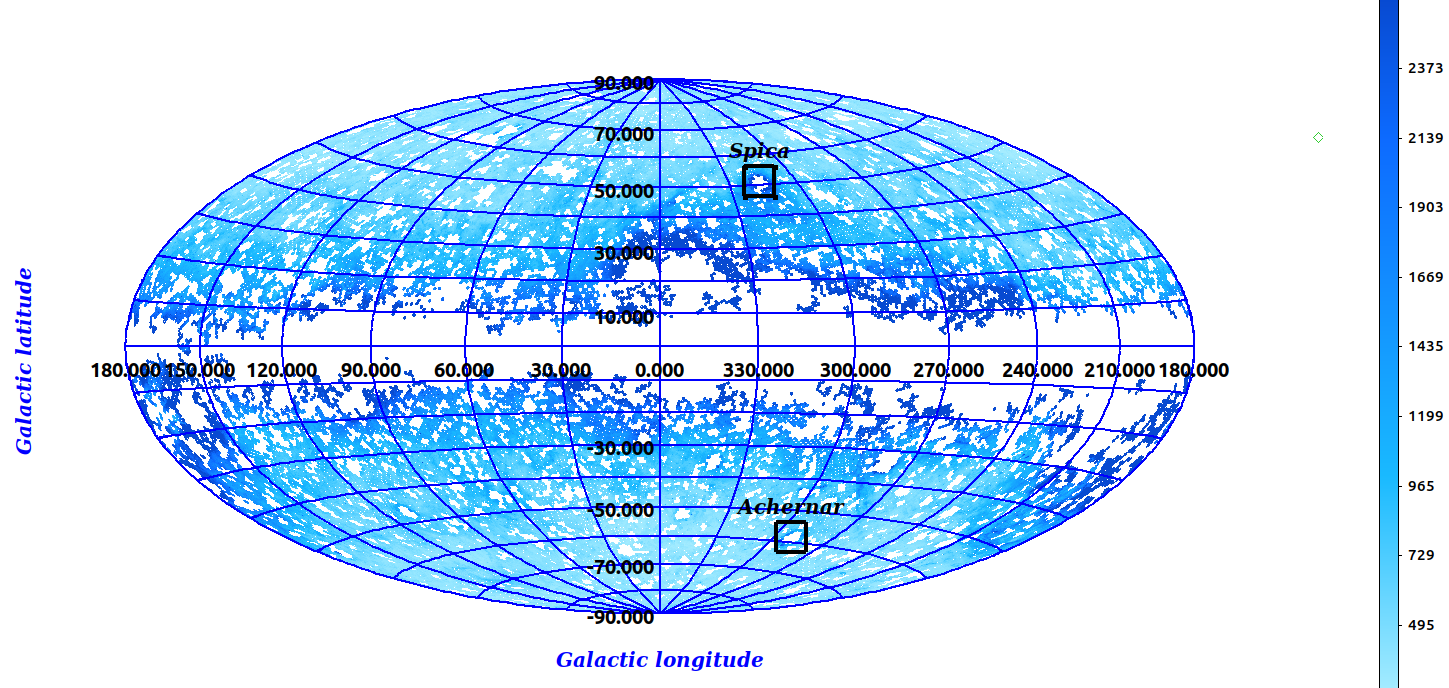}
    \caption{Diffuse sky image of FUV halos around Spica and Achernar observed by GALEX , \textit{Credit}: \citet{Murthy2011_UV_halos}. }
    \label{fig:diffuse-sky-map}
\end{figure*}

\subsubsection{Achernar}
\vspace{0.1cm}
Achernar (l = 290.84\degree , b = -58.79\degree) also known as $\alpha$ Eridani is a main-sequence variable star having a spectral classification as B3 Vpe (\citet{Murthy2011_UV_halos}). It is the 9th brightest star having a visible magnitude of 0.46 which is located in the constellation of Eridanus. From the variation in polarization levels, \citet{Carciofi2007} had shown the existence of  circumstellar disk around Achernar which is formed due to several mass ejection episodes of circumstellar materials. 
\citet{Kervella2008} found that there exists a polar envelope due to the fast winds ejected from the polar caps of Achernar and the presence of this envelope is detected by IR excess emission. 
Achernar is situated inside the Local Bubble as shown in fig \ref{fig:loop-bubble} and is relatively closer to us than Spica.  

The observed FUV intensities in the halo of Achernar are modeled using a similar double layered plane-parallel sheet as for Spica. Since the total amount of material in front of Achernar is unknown, we have considered two values for the optical depth. In the first case, we assume the same optical depth ($\tau$=0.1) as for Spica and the dust is distributed among the two sheets such that 20\% of the total matter is in the first sheet and remaining 80\% in the second sheet. Secondly, we consider 10\% and 90\% of the material in the two sheets respectively. The distance of the first scattering layer ($dl_{1}$) from Achernar, is varied from 1 to 10 pc to obtain the best-fit distance for the first sheet and the second sheet is placed at a distance of 30 PC ($dl_{2}$) for providing a 20-30\% offset to the modeled intensities.
Further, we also consider a tau value scaled by best-fit intensities obtained using Draine's phase function and consider the two cases of distributing the material in two sheets with two different proportions as done for $\tau$ = 0.1.
\section{Results}
The FUV intensity intensities of the halos are governed by parameters such as albedo and phase function of the grains, density of grains in the scattering layer, star's intrinsic luminosity and the star-dust geometry. By utilizing theoretical phase functions corresponding to different types of grains, we have derived the best-fit grain models and the corresponding phase functions responsible for the halos of Spica and Achernar. 
\subsection{Best-fit model for Spica}

\subsubsection*{\textbf{a. Astronomical Silicate and Graphite}}

\begin{figure}[H]
\centering
\includegraphics[height=8.0cm,width=8.0cm]
{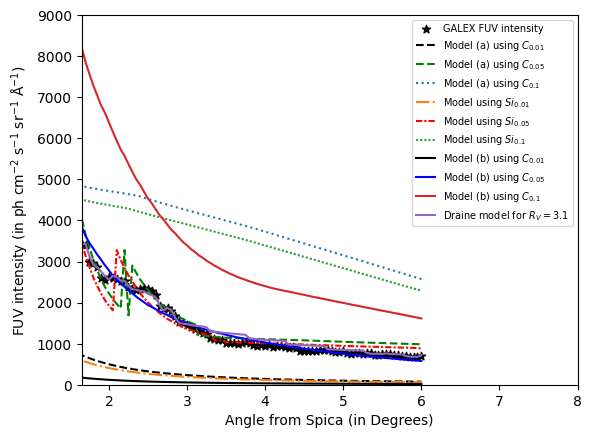}
    \caption{Variation of observed and best-fit model intensities in FUV (obtained using porous aggregates made up of graphite and silicates and Draine's model for $R_V$ = 3.1) with the angular separation (in degrees) from Spica. minimum $\chi_{\nu}^2$= 0.803 for $C_{0.05}$ model (a), minimum $\chi_{\nu}^2$= 0.591 for $Si_{0.05}$ model, minimum $\chi_{\nu}^2$= 0.294 for $C_{0.05}$ model (b) and minimum $\chi_{\nu}^2$= 0.123 for Draine's model for $R_V$ = 3.1}
    \label{fig:intensity-vs-angle-6-porous-aggregates-Spica}
\end{figure}

The different model intensities along with observed intensities are plotted as a function of the angular separation from the star in fig \ref{fig:intensity-vs-angle-6-porous-aggregates-Spica}. It is observed that the model intensity increases on increasing the size of porous aggregate. The minimum $\chi_{\nu}^2$ values ($\nu$ = 87) for each model are presented in Table \ref{model-Spica}. It is evident that the graphite grains model (a) with effective radius, 0.05$\mu$m provides the best fit with a minimum $\chi_{\nu}^2$value of 0.804 compared to the other porous aggregates. The Silicate model with a similar grain size gives a lower minimum $\chi_{\nu}^2$value of 0.591. 
Draine's model for R$_{V}$ is a poor fit with a minimum $\chi_{\nu}^2$value of 0.122.
\begin{table}[H]
\caption{Best-fit model parameters for Spica}
\vspace{0.2cm}
\resizebox{0.5\textwidth}{!}{%
\begin{tabular}{|c|c|c|c|c|}
\hline
\textbf{Composition} & \textbf{Model}  & \textbf{Albedo} & \textbf{Asymmetry} & \textbf{minimum $\chi_{\nu}^2$  } \\
&\textbf{}&&\textbf{parameter} & \\
&& \textbf{($a$)}& \textbf{($g$)}& \textbf{($\nu$ = 87)}\\
\hline
Silicate&$Si_{ 0.05}$& 0.530 & 0.854 & 0.591  \\
\hline
Graphite(a) &$C_{ 0.05}$& 0.598 & 0.853 & 0.804   \\
\hline
 Graphite(b) &$C_{ 0.05}$& 0.253 & 0.859 & 0.294  \\
\hline
Gr+Sil&Draine2003a&0.41986  &0.66485 & 0.122 \\
\hline
\end{tabular}
}
\label{model-Spica}
\end{table}

\begin{figure}[H]
\centering
\includegraphics[height=7.0cm,width=8.0cm]  {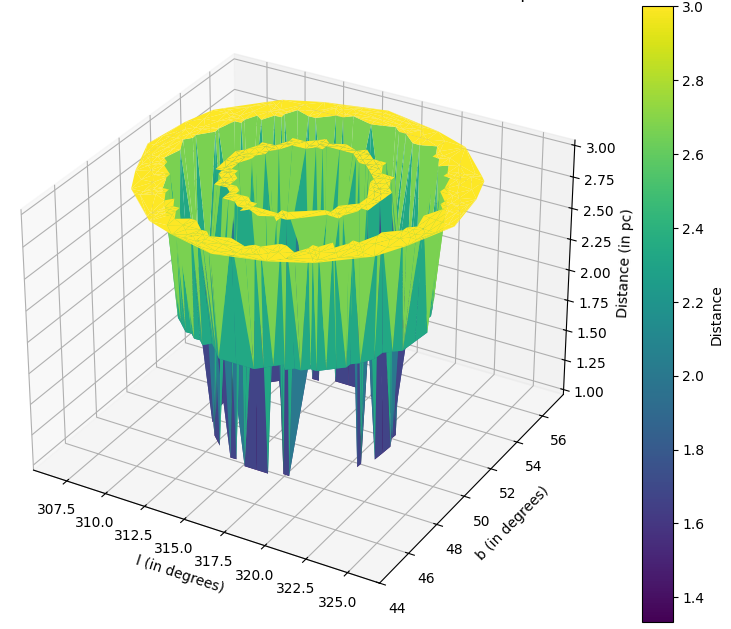}
\caption{Variation of the dust cloud distance from Spica (in pc) for $C_{0.05}$ model (a) plotted as a function of location (in galactic coordinates, in degrees) in the halo.}
\label{fig:C-0.05-distance-vs-angle-from-Spica-model-a}
\end{figure}

The dust cloud distance from Spica ($dl_{1}$) is constrained between 1-3 pc by using astronomical graphite grain model (a) with a effective radius of 0.05 $\mu$m as shown in fig. \ref{fig:C-0.05-distance-vs-angle-from-Spica-model-a}. 

Further, we compare the observed and model intensities for the best-fit model. 
We find that there is a reasonable linear fit between the two intensities and the correlation co-efficient is 0.9514.

\subsubsection*{\textbf{b. Amorphous Carbon and Silicate}}
On considering the porous aggregate model based on amorphous carbon and silicate, the best fit model is achieved for Spica using $SiE_{0.05}$ i.e. Enstatite (amorphous silicate) model with a minimum $\chi^2_\nu$ value of 0.943 as compared to other models. $aC_{0.05}$ provides a  comparatively lower value of minimum $\chi^2_\nu$ as 0.235 and $SiF_{0.05}$ i.e. Forsterite (amorphous silicate) model provides a minimum $\chi^2_\nu$ of 0.651.
\begin{table}[H]
\textbf{\caption{Model parameters for Spica}}
\vspace{0.2cm}
\resizebox{0.5\textwidth}{!}{%
\begin{tabular}{|c|c|c|c|c|}
\hline
\textbf{Composition} & \textbf{Model}  & \textbf{Albedo} & \textbf{Asymmetry} & \textbf{Minimum $\chi_{\nu}^2$ } \\
& & &\textbf{parameter} & \textbf{Spica} \\
&& \textbf{($a$)}& \textbf{($g$)}& \textbf{($\nu$ = 87)}\\

\hline
\multirow{3}{*}{}{Amorphous} & \multirow{3}{*}{}{$aC_{ 0.01}$} & 0.056 & 0.079 & 22.515 \\
\cline{2-5}
\multirow{3}{*}{}{Carbon}  & \multirow{3}{*}{}{$aC_{ 0.05}$} & 0.416 & 0.857  & 0.235 \\
\cline{2-5}
 & \multirow{3}{*}{}{$aC_{ 0.1}$} & 0.557 & 0.926 & 33.390 \\
 
\hline
\multirow{3}{*}{}{Amorphous} & \multirow{3}{*}{}{$SiF_{ 0.01}$} & 0.077 & 0.079  & 20.898 \\
\cline{2-5}
\multirow{3}{*}{}{Silicate} & \multirow{3}{*}{}{$SiF_{ 0.05}$} & 0.492 & 0.855 & 0.651\\
\cline{2-5}
\multirow{3}{*}{}{(Forsterite)}  & \multirow{3}{*}{}{$SiF_{ 0.1}$} & 0.612 & 0.917 & 44.213\\

\hline
\multirow{3}{*}{}{Amorphous} & \multirow{3}{*}{}{$SiE_{ 0.01}$} & 0.121 & 0.079   & 17.694\\
\cline{2-5}
\multirow{3}{*}{}{Silicate} & \multirow{3}{*}{}{$SiE_{ 0.05}$} & 0.613 & 0.850  & 0.943\\
\cline{2-5}
\multirow{3}{*}{}{(Enstatite)}  & \multirow{3}{*}{}{$SiE_{ 0.1}$} & 0.703 & 0.902 & 61.437 \\
\hline
\end{tabular}
}
\label{model-Spica}
\end{table}

\begin{figure}[H]
\centering
\includegraphics[height=6.5cm,width=8.0cm]
{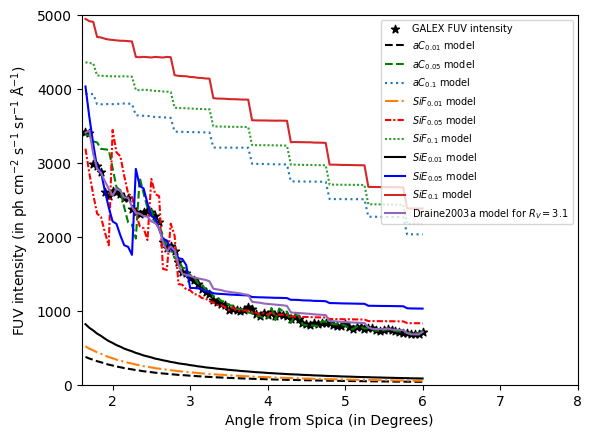}
    \caption{Variation of observed and best-fit model intensities in FUV (obtained using porous aggregates made up of amorphous carbon, amorphous silicate (amorphous enstatite and amorphous forsterite ) and Draine's grain model for $R_V$ = 3.1) with the angular separation (in degrees) from Spica.} 
    
    \label{fig:intensity-vs-angle-new-grain-models-Spica}
\end{figure}

\begin{figure}[H]
\centering
\includegraphics[height=7.0cm,width=8.0cm]  {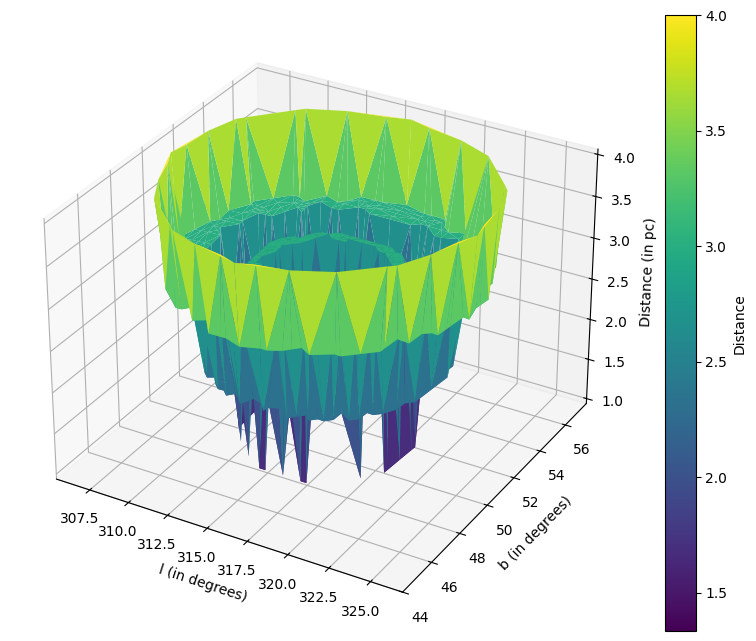}
\caption{Variation of the dust cloud distance from Spica (in pc) for $SiE_{0.05}$ model plotted as a function of location (in galactic coordinates, in degrees) in the halo.}
\label{fig:SiE-0.05-distance-vs-gal-cord-for-Spica}
\end{figure}

The dust cloud distance from Spica ($dl_1$) is constrained between 1-4 pc by using amorphous silicate based grain model ($SiE_{0.05}$) 
with a effective radius of 0.05 $\mu$m as shown in fig. \ref{fig:SiE-0.05-distance-vs-gal-cord-for-Spica}.

\subsection{Best-fit model for Achernar}

\subsubsection*{\textbf{a. Astronomical Silicate and Graphite}}
\label{sec:Astr-Sil-Gra}

\begin{figure}[H]
\centering
\includegraphics[height=6.0cm,width=8.0cm]
{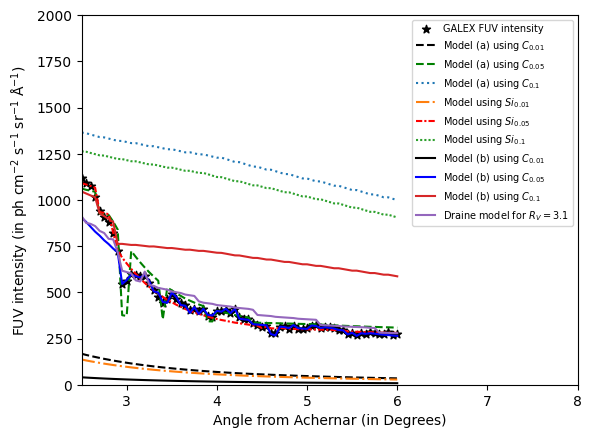}
    \caption{Variation of observed and best-fit model intensities in FUV for Case 3 (using model (a) and model (b) for graphite, model for silicates and Draine's model for $R_V$ = 3.1) with the angular separation (in degrees) from Achernar.
    minimum $\chi_{\nu}^2$= 1.098 for $C_{0.05}$ model (a), minimum $\chi_{\nu}^2$= 0.390 for $Si_{0.05}$ model, minimum $\chi_{\nu}^2$= 1.341 for $C_{0.05}$ model (b) and minimum $\chi_{\nu}^2$= 1.81 for Draine's model ($R_V$ = 3.1) }
    
  \label{fig:method-3-model-intensity-vs-angle-6-porous-aggregates-Achernar_20}
\end{figure}

\begin{figure}[H]
\centering
\includegraphics[height=6.0cm,width=8.0cm]
{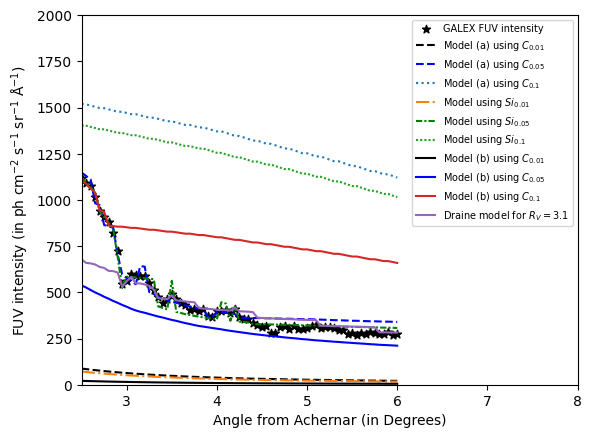}
    \caption{Variation of observed and best-fit model intensities in FUV for Case 4 (using model (a) and model (b) for graphite, model for silicates and Draine's model for $R_V$ = 3.1) with the angular separation (in degrees) from Achernar. minimum $\chi_{\nu}^2$= 0.760 for $C_{0.05}$ model (a), minimum $\chi_{\nu}^2$= 0.372 for $Si_{0.05}$ model and minimum $\chi_{\nu}^2$= 0.038 for Draine's model ($R_V$ = 3.1).}
    
  \label{fig:method-4-model-intensity-vs-angle-6-porous-aggregates-Achernar_10}
\end{figure}

\begin{table}[H]
    \caption{Best-fit model parameters for Achernar}
\resizebox{0.5\textwidth}{!}{%
    \begin{tabular}{|c|c|c|c|c|c|}
    \hline
    \textbf{Case} & \textbf{Composition} & \textbf{Model}  & \textbf{Albedo} & \textbf{Asymmetry} & \textbf{minimum $\chi_{\nu}^2$  } \\
    & & \textbf{}&&\textbf{parameter} & \\
    && &\textbf{($a$)}& \textbf{($g$)}& \textbf{($\nu$ = 70)}\\
    \hline
    \multirow{2}{*}{}{1} & Graphite (b) & $C_{0.05}$ & 0.253 & 0.859 & 7.078 \\
    \cline{2-6}
    & Gr+Sil & Draine2003a &0.41986 & 0.66485 & 1.806\\
    \hline
    \multirow{2}{*}{}{2} & Graphite (b) & $C_{0.05}$ & 0.253 & 0.859 & 6.192 \\
    \cline{2-6}
    & Gr+Sil & Draine2003a &0.41986 & 0.66485 & 279.91\\
    \hline
    \multirow{4}{*}{}{3} & Graphite (a) & $C_{0.05}$ & 0.598 & 0.853 & 1.098 \\
    \cline{2-6}
    &Silicate & $Si_{0.05}$ &0.530 & 0.854 & 0.390 \\
    \cline{2-6}
    &Graphite (b)& $C_{0.05}$ & 0.253 & 0.859 & 1.341\\
    \cline{2-6}
    & Gr+Sil & Draine2003a &0.41986 & 0.66485 & 1.81\\
    \hline
    \multirow{3}{*}{}{4} & Graphite (a) & $C_{0.05}$ & 0.598 & 0.853 & 0.760 \\
    \cline{2-6}
    & Silicate & $Si_{0.05}$ &0.530 & 0.854 & 0.372 \\
    \cline{2-6}
    & Gr+Sil & Draine2003a &0.41986 & 0.66485 & 6.038\\
    \hline
    \end{tabular}
    }

\label{tab:best_fit_model_parameters_for_Achernar}
\end{table}
The best-fit model intensities for Achernar is achieved by making trials conducted for 4 different cases as being mentioned below:
\begin{itemize}
\item 
\textbf{Case 1 : $\tau_{tot}$=0.1, $\tau_{1}=0.02$, $\tau_{2}=0.08$}\\
By considering the total optical depth ( $\tau_{tot}$) to be distributed among the two sheets such that 20 \% ($\tau_{1}$) is in the first sheet and remaining 80 \% ($\tau_{2}$) in the second sheet, we find that the model intensities obtained for different porous grains do not give a good fit. However, Draine's Rv=3.1 model gives a comparatively better fit with a minimum $\chi_{\nu}^2$ of 1.806 .
\vspace{0.1cm}
\item
\textbf{Case 2 : $\tau_{tot}$=0.1, $\tau_{1}=0.01$, $\tau_{2}=0.09$}\\
By considering  $\tau_{tot}$ to be distributed among the two sheets  (10 \% and 90 \% of the total $\tau$),  for different grain models we do not get a reasonable fit.\\
Since, the two distributions considered so far (Case 1,2) do not provide good fits, we scale the optical depth value based on the ratio of observed to model intensities obtained using Draine's model, resulting in a lower total optical depth of 0.032 for Achernar. Once again, we consider dividing the material among the two sheets in different proportions as given below. 
\vspace{0.1cm}
\item
\textbf{Case 3 : $\tau_{tot}$=0.032, $\tau_{1}=0.006$, $\tau_{2}=0.026$}\\
The material is divided among two sheets so that 20\% of the material is in the first sheet and the remaining material in the second sheet. Fig. \ref{fig:method-3-model-intensity-vs-angle-6-porous-aggregates-Achernar_20} shows the model intensities in this case as a function of angular separation from the star.
Based on the minimum $\chi_{\nu}^2$  values, we find that both the graphite grain models (a \& b) with grain size of 0.05 $\mu$m give good fits with minimum $\chi_{\nu}^2$  values of 1.098 and 1.341 respectively. Draine's model with Rv=3.1 also gives a moderate fit to the observed intensities with a minimum $\chi_{\nu}^2$  of 1.81.
Further, we have compared the observed and model intensities for best-fit model obtained in \textbf{case 3} for $C_{0.05}$ Model (a) and found that there is a reasonable linear fit between these two intensities with a correlation co-efficient of 0.975 . 
\vspace{0.1cm}
\item
\textbf{Case 4 : $\tau_{tot}$=0.032, $\tau_{1}=0.003$, $\tau_{2}=0.029$}\\
The material is divided among two sheets so that 10\% of the material is in the first sheet and the remaining material in the second sheet. Fig. \ref{fig:method-4-model-intensity-vs-angle-6-porous-aggregates-Achernar_10} shows the model intensities in this case as a function of angular separation from the star.
We find that, the graphite model (a) with 0.05$\mu$m sized grains provides a moderately improved fit with a minimum $\chi_{\nu}^2$ = 0.76, compared to other models.
\end{itemize}
\vspace{-0.2cm}

\begin{figure}[H]
\centering
\includegraphics[height=6.5cm,width=8.0cm]  {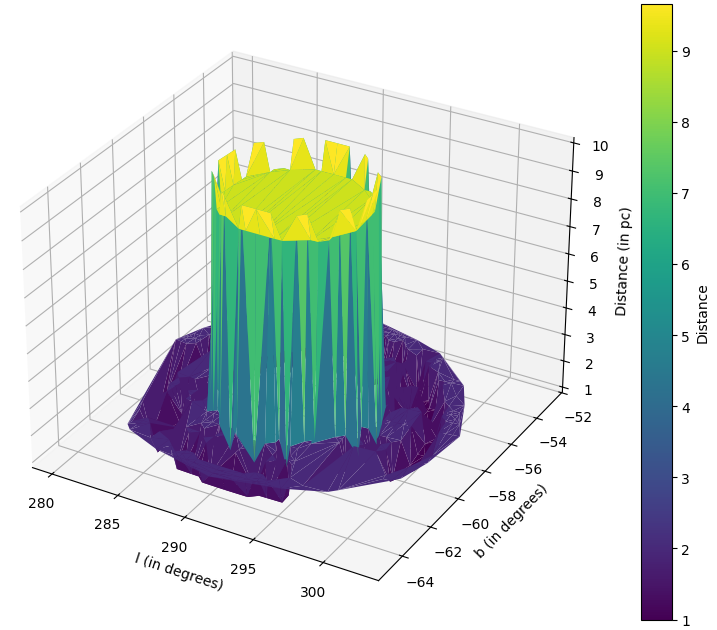}
\caption{Variation of the dust cloud distance from Achernar (in pc) for case 3 of $C_{0.05}$ model (a) plotted as a function of location (in galactic coordinates, in degrees) in the halo.}
\label{fig:C-0.05-distance-vs-angle-from-Achernar-model-a}
\end{figure}

On considering the best-fit model obtained for Achernar in case 3 for astronomical graphite grain ($C_{0.05}$ model (a) ) having an effective radius of 0.05 $\mu$m, the dust cloud distance from Achernar ($dl_{1}$), varies between 1 to 10 pc as shown in fig \ref{fig:C-0.05-distance-vs-angle-from-Achernar-model-a}, with most of the locations within 2 pc from the star.


\subsubsection*{\textbf{b. Amorphous Carbon and  Silicate}}
From the results obtained using astronomical silicate and graphite for Achernar, we have considered case 3 and 4 as an approach towards obtaining a best fit model for Achernar using amorphous carbon and amorphous silicate as shown in Table \ref{tab:newmodel-Achernar}. The best fit model is obtained for amorphous carbon ($aC_{0.05}$) with a minimum $\chi_{\nu}^2$ of 1.05 by considering case 3. The amorphous silicate model in case 3 has provided a minimum $\chi_{\nu}^2$ of 1.214 for $SiE_{0.05}$ (Enstatite). On considering case 4, a minimum $\chi_{\nu}^2$ of 1.509 is achieved for amorphous carbon ($aC_{0.05}$) model and amorphous silicate model provides a minimum $\chi_{\nu}^2$ of 1.208 for $SiE_{0.05}$.

\begin{table}[H]
\caption{Model parameters for Achernar}
\vspace{0.2cm}
\resizebox{0.5\textwidth}{!}{%
\begin{tabular}{|c|c|c|c|c|c|}
\hline
\textbf{Composition} & \textbf{Model}  & \textbf{Albedo} & \textbf{Asymmetry} & \multicolumn{2}{|c|}{\textbf{Minimum $\chi_{\nu}^2$  }} \\
\cline{5-6}
& & &\textbf{parameter} & \multicolumn{2}{|c|}{\textbf{Achernar ($\nu$ = 70)}}\\
\cline{5-6}
&& \textbf{($a$)}& \textbf{($g$)} & \textbf{Case 3} & \textbf{Case 4}\\

\hline
\multirow{3}{*}{}{Amorphous} & \multirow{3}{*}{}{$aC_{ 0.01}$} & 0.056 & 0.079 & 83.089 & 90.390 \\
\cline{2-6}
\multirow{3}{*}{}{Carbon}  & \multirow{3}{*}{}{$aC_{ 0.05}$} & 0.416 & 0.857 & 1.050 & 1.509\\
\cline{2-6}
 & \multirow{3}{*}{}{$aC_{ 0.1}$} & 0.557 & 0.926  & 116.508 & 169.379\\
\hline

\multirow{3}{*}{}{Amorphous} & \multirow{3}{*}{}{$SiF_{ 0.01}$} & 0.077 & 0.079  & 77.310 & 87.097\\
\cline{2-6}
\multirow{3}{*}{}{Silicate} & \multirow{3}{*}{}{$SiF_{ 0.05}$} & 0.492 & 0.855  & 0.692 & 0.372\\
\cline{2-6}
\multirow{3}{*}{}{(Forsterite)}  & \multirow{3}{*}{}{$SiF_{ 0.1}$} & 0.612 & 0.917  & 155.273 & 221.235\\
\hline

\multirow{3}{*}{}{Amorphous} & \multirow{3}{*}{}{$SiE_{ 0.01}$} & 0.121 & 0.079  & 65.831 & 80.362\\
\cline{2-6}
\multirow{3}{*}{}{Silicate} & \multirow{3}{*}{}{$SiE_{ 0.05}$} & 0.613 & 0.850  & 1.214 & 1.208\\
\cline{2-6}
\multirow{3}{*}{}{(Enstatite)}  & \multirow{3}{*}{}{$SiE_{ 0.1}$} & 0.703 & 0.902  & 216.963 & 301.056 \\
\hline
\end{tabular}
}
\label{tab:newmodel-Achernar}
\end{table}

\begin{figure}[H]
\centering
\includegraphics[height=6.0cm,width=8.0cm]
{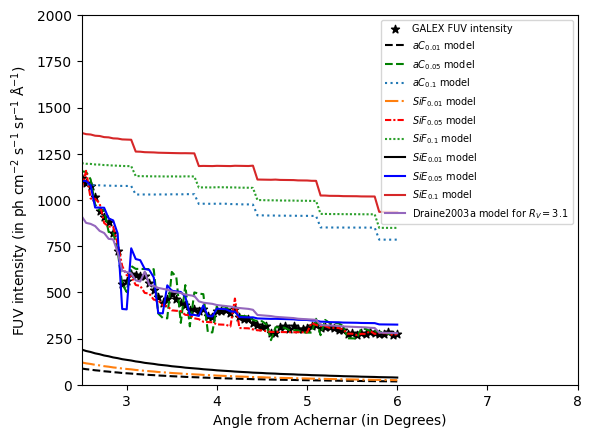}
    \caption{Variation of observed and best-fit model intensities in FUV for Case 3 using porous grain model (consisting of carbon and silicate in amorphous form) and Draine's grain model for $R_V$ = 3.1 with the angular separation (in degrees) from Achernar.}
    
  \label{fig:method-3-model-intensity-vs-angle-new-grain-models-Achernar_20}
\end{figure}

\begin{figure}[H]
\centering
\includegraphics[height=6.0cm,width=8.0cm]
{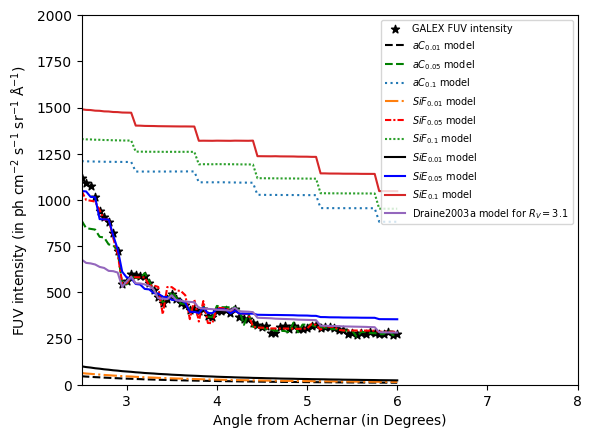}
    \caption{Variation of observed and best-fit model intensities in FUV for Case 4 (using porous grain model (consisting of carbon and silicate in amorphous form) and Draine's grain model for $R_V$ = 3.1 with the angular separation (in degrees) from Achernar.}
  \label{fig:method-4-model-intensity-vs-angle-new-grain-models--Achernar_10}
\end{figure}

\begin{figure}[H]
\centering
\includegraphics[height=8.0cm,width=8.0cm]{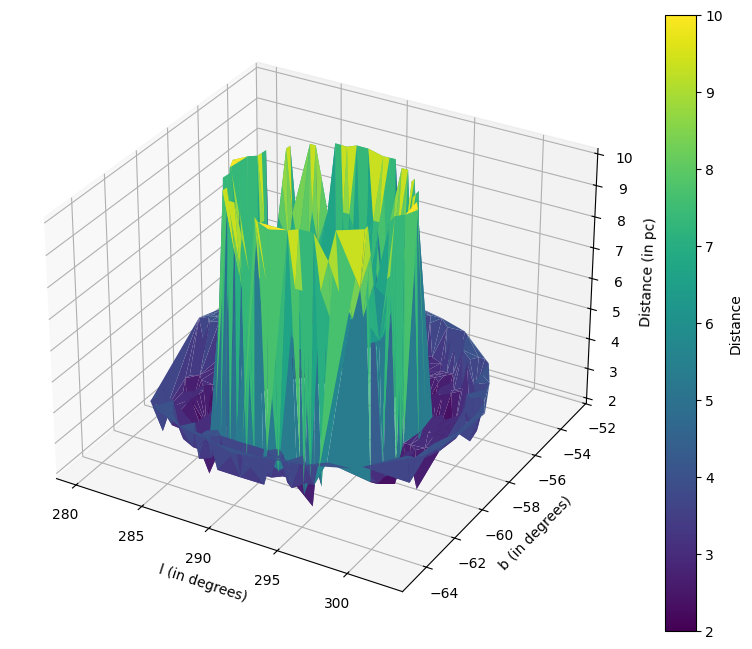}
\caption{Variation of the dust cloud distance from Achernar (in pc) for case 3 of $C_{0.05}$ model plotted as a function of location (in galactic coordinates, in degrees) in the halo.}
\label{Achernarcase3}
\end{figure}

The dust cloud distance from Achernar ($dl_1$) varies between 1 to 10 pc but mostly it is constrained between 2-4 pc by using amorphous carbon based grain model ($C_{0.05}$) 
with a effective radius of 0.05 $\mu$m as shown in fig. \ref{Achernarcase3}.

\subsection{Comparison of best-fit model phase function with other phase functions}
\begin{figure}[H]
\centering
\includegraphics[height=6.5cm,width=8.0cm]{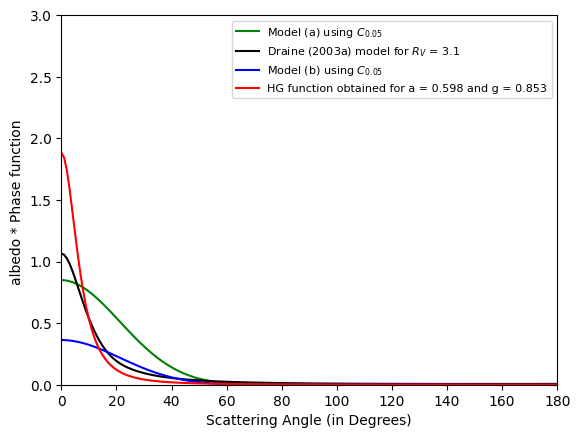}
\caption{Comparison of phase functions of model (a) and model (b) $C_{0.05}$ for Spica and Achernar with HG and Draine's R$_{V}$=3.1 phase functions.}
\label{fig:oldmodelphi}
\end{figure}

\begin{figure}[H]
\centering
\includegraphics[height=6.5cm,width=8.0cm]{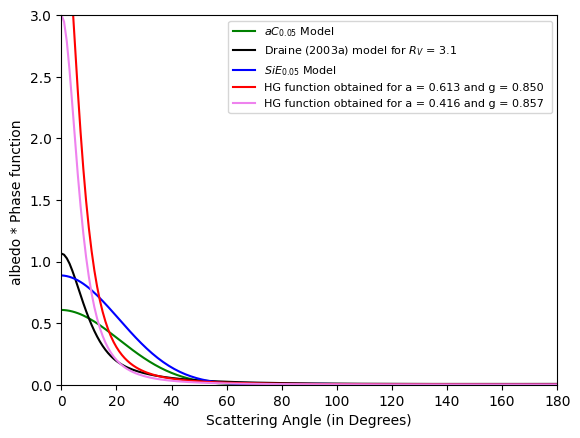}
\caption{Comparison of phase functions of $SiE_{0.05}$ and $aC_{0.05}$ model for Spica and Achernar respectively with the Draine's R$_{V}$=3.1 and best-fit HG phase functions.}
\label{fig:newmodelphi}
\end{figure}

We have compared the best-fit scattering phase function obtained for Spica and Achernar with the HG phase function computed using our best-fit g obtained for the two stars and phase function provided by \citet{Draine2003_Scattering} for $R_V$= 3.1 in figs. \ref{fig:oldmodelphi} and \ref{fig:newmodelphi} respectively. Since the single scattering intensities linearly scale with the value of $a$, we have plotted the product of the phase function with the corresponding $a$ values given in table \ref{tab:parameters-comparison-of-phi}. It is evident that the phase functions deviate considerably from each other at smaller scattering angles. The porous aggregates made up of astronomical graphite and amorphous silicate grains show comparatively larger amount of scattering than other grains and phase functions for these angles. From fig. \ref{fig:newmodelphi} it is observed that HG phase function doesn't vary much for both amorphous carbon and silicate grains. It is also evident from figs. \ref{fig:oldmodelphi} and \ref{fig:newmodelphi}, that the HG phase function shows a larger deviation from the theoretical phase function at smaller angles. Hence, it is shown that HG phase function cannot adequately explain the dust scattering at smaller angles.

\begin{table}[H]
\caption{Parameters involved for comparing the phase functions}
\vspace{0.2cm}
\resizebox{0.5\textwidth}{!}{%
\begin{tabular}{|c|c|c|c|}
    \hline
    \textbf{Phase functions} & \textbf{Grain Type}  & \textbf{a} & \textbf{g} \\ 
    & &  & \\
    \hline
     \multirow{3}{*}{} {HG} & Graphite (a)  & 0.598 & 0.853\\
     \cline{2-4}
     & Amorphous Carbon & 0.416 & 0.857 \\
     \cline{2-4}
     & Enstatite & 0.613 & 0.850 \\
    \hline
    
     $D2003a^{\textcolor{blue}{1}}$ & Gr+Sil  & 0.41986 & 0.66485\\
    \hline
    \end{tabular}
    }
    \label{tab:parameters-comparison-of-phi}
\end{table}

\textbf{Note:}
$^{\textcolor{blue}{1}}$\citet{Draine2003_Scattering} ($R_V = 3.1$).

\subsection{Comparison with earlier studies}
\begin{table}[H]
\caption{Dust scattering model comparison for Spica}
\vspace{0.2cm}
\resizebox{0.5\textwidth}{!}{%
    \begin{tabular}{|c|c|c|c|c|c|}
    \hline
    \textbf{Parameters} & \textbf{This work} & \textbf{This Work} & \textbf{This Work} &\textbf{S2013$^{\textcolor{blue}{a}}$} & \textbf{M2011$^{\textcolor{blue}{b}}$}\\ 
    \hline
    Phase function& C 0.05a & \textbf{$SiE_{0.05}$}& $D2003a^{\textcolor{blue}{1}}$ & HG & HG\\
    \hline
    a & 0.598 & 0.613 &0.41986 & 0.26 $\pm$ 0.1 & 0.5 \\
    \hline
    g & 0.853 & 0.850 &0.66485 & 0.58 $\pm$ 0.11 & 0.6 \\
    \hline
    $dl_1$ & 2-3 pc & 1-4 pc &1-10 pc & 3.65 $\pm$ 1.05 pc & 2 pc \\
    \hline
    minimum $\chi_{\nu}^2$& 0.804 & 0.943 & 0.122 & 0.438 & 1.34\\
    \hline
    \end{tabular} 
    }
    \label{tab:dust-scattering-parameters-comparison}
\end{table}

\begin{table}[H]
\caption{Dust scattering model comparison for Achernar}
\vspace{0.2cm}
\resizebox{0.5\textwidth}{!}{%
    \begin{tabular}{|c|c|c|c|c|}
    \hline
    \textbf{Parameters} & \textbf{This work} & \textbf{This Work} & \textbf{This Work} &\textbf{M2011$^{\textcolor{blue}{b}}$}\\ 
    \hline
    Phase function& C 0.05a & \textbf{$aC_{0.05}$}& $D2003a^{\textcolor{blue}{1}}$ & HG\\
    \hline
    a & 0.598 & 0.416 &0.41986  & 0.11 $\pm$ 0.03\\
    \hline
    g & 0.853 & 0.857 &0.66485  & 0.75 $\pm$ 0.05 \\
    \hline
    $dl_1$ & 1-2 pc  & 2-4 pc & 1-4 pc & 3 pc  \\
    \hline
    minimum $\chi_{\nu}^2$& 1.098 & 1.05 & 1.810 & 0.915 \\
    \hline
    \end{tabular}
    }
    \label{tab:dust-scattering-parameters-comparison-Achernar}
\end{table}

\textbf{Notes:} \\
$^{\textcolor{blue}{a}}$ \citet{Shalima2013}
\\ $^{\textcolor{blue}{b}}$ \citet{Murthy2011_UV_halos}\\
$^{\textcolor{blue}{1}}$ \citet{Draine2003_Scattering} ($R_V = 3.1$)
\\Tables \ref{tab:dust-scattering-parameters-comparison} and \ref{tab:dust-scattering-parameters-comparison-Achernar} show the comparison between the dust parameters of the best-fit porous aggregate model, Draine's Rv=3.1 model, and the earlier results of \citet{Murthy2011_UV_halos} and \citet{Shalima2013} using the HG phase function for grains towards Spica and Achernar. We find that the obtained albedo value for Spica is in agreement with the results of \citet{Murthy2011_UV_halos}, whereas the value for Achernar is higher. The $g$ value obtained here is larger than the values obtained by \citet{Murthy2011_UV_halos} for both stars and \citet{Shalima2013} for Spica.


We find that the $a$ and $g$ values obtained here are closer to the results ($a$ = 0.5 and $g$ = 0.9) obtained by \citet{Witt1994} for Diffuse Galactic Light (DGL).

In the current study, the structure of the dust grain is considered to be that of porous aggregates. But the ISM dust grains can always have other  morphologies such as core-mantle grains, agglomerated debris, ellipsoids, rough fractal aggregates etc. \citep{2006Zubko,2017Jones,2022Halder}. It is therefore important to incorporate them for impovizing the model. Though we have compared our results with Draine's model which considers a size distribution for spherical grains, such an approach for aggregate grains has not been included here and will be incorporated in a forthcoming paper.


\section{Conclusions}
Using the porous aggregates for modeling the single scattering intensities, we have constrained the phase function of the dust grains responsible for the UV halo around Spica and Achernar. From the results obtained here, we arrive at the following conclusions:
\begin{itemize}
\item 
Both carbonaceous dust aggregates and amorphous silicates with typical sizes of 0.05$\mu$m are the most likely constituents of the scattering dust grains towards Spica. On the other hand, carbonaceous dust aggregates with typical sizes of 0.05$\mu$m are the most likely constituents of the scattering dust grains towards Achernar.
\vspace{-0.1cm}
\item 
For Spica, the scattering phase function of the best-fit porous aggregate model for $SiE_{0.05}$ corresponds to $a$ and $g$ values of 0.613 and 0.85 respectively. For Achernar, the scattering phase function of the best-fit model for $aC_{0.05}$ and $C_{0.05}$ corresponds to albedo values 0.416, 0.598 and asymmetric parameter values 0.857, 0.853 respectively. Both the cases indicates highly forward-scattering grains.  
\vspace{-0.1cm}
\item
The dust cloud having a contribution of $\sim$ 20\% towards the FUV scattering intensities is located between 1-4 pc from Spica, in agreement with earlier results.
\vspace{-0.1cm}
\item
The medium in front of Achernar has a low optical depth ($\tau$) of 0.032 compared to Spica ($\tau$ as 0.1) and lies mostly mostly between 2-4 pc from Achernar with a contribution of $\sim$ 20\% towards the FUV scattering intensities, except at small angular separation, where it extends to 10 pc from the star. 
\vspace{-0.1cm}
\item
A comparison with previous studies shows that $a$ of our best-fit grain model is in good agreement with the results obtained by \citet{Murthy2011_UV_halos} for Spica, but more than $\sim$ 5 times higher for Achernar. 
\vspace{-0.1cm}
\item 
The $g$ value corresponding to the best-fit porous aggregate phase function for Achernar is comparatively higher than the values obtained by \citet{Murthy2011_UV_halos} and corresponding to the R$_{V}$ =3.1 grain model of \citet{Draine2003_Scattering}. 
\vspace{-0.1cm}
\item We could constrain the dust cloud distance to within \textbf{1-4} pc for majority of the locations in the halo of Achernar, using the best-fit porous aggregate model which is in agreement with the results for the Draine's model \citep{Draine2003_Scattering}, which puts the dust cloud at a distance of 4 pc. 
\end{itemize}

However, it is important to consider other grain morphologies with the size distribution in order to reduce the uncertainties in the dust grain properties of FUV halos.

\paragraph{Data Availability Statement}
The data used here in this work is taken from the NASA’s GALEX mission observation data provided by \citet{Murthy2011_UV_halos}.

\paragraph{Author Contributions}
PS proposed the concept and JM provided the inputs for the data analysis utilized in this work.
KU computed the porous aggregate phase functions using DDSCAT. NDC did the modeling calculations for the FUV halos and further  created the plots.
 All authors have contributed for preparing the manuscript.

\paragraph{Funding} No funding is associated with the present work.

\section*{Acknowledgements}
PS, KU \& NDC sincerely thank MCNS, MAHE for providing the facilities and support. KU would like to thank MAHE for providing TMA Pai Scholarship during this work. We would like to thank the anonymous referee for the useful suggestions which immensely improved the quality of the work.

\appendix

\section{Extinction data for stars nearby Spica}
\vspace{-0.09cm}
\begin{table}[H]
\caption{Extinction parameter for Stars selected from \citet{Gontcharov2019} near Spica}

\resizebox{0.4\textwidth}{!}{%
    \begin{tabular}{|c|c|c|c|c|}
    \hline
    \textbf{Star} & \textbf{l ($\degree$)} & \textbf{b ($\degree$)} & \textbf{Distance (pc) }& \textbf{$R_V$} \\
    \hline
    *iot Vir & 337.7416 & 51.0653 & 22 & 3.16 \\
    HD 124106 & 331.6227 & 45.7369 & 23 & 3.16 \\
    *zet Vir & 325.2438 & 60.389 & 22.5 & 3.18 \\
    HD 129923 &	342.4082 & 42.5269 & 190.3 & 3.19 \\
    *66 Vir & 	318.0872 & 56.7299 & 31.1 & 3.2 \\
    HD 119638 & 321.934	& 46.671 & 33.8 & 3.2 \\
    *38 Vir & 303.7844 & 59.3158 & 33.2 & 3.21\\
    HD 126679 & 334.8509 & 42.0114 & 45.5 & 3.23\\
    HD 113022 & 317.0155 & 80.9923 & 40.5 & 3.26\\
    HD 125612 & 331.3895 & 40.3954 & 57.6 & 3.27\\
    HD 116259 & 313.334	& 44.208 & 59.3 & 3.28 \\
    HD 121301 & 324.7158 &44.6384 & 61.5 & 3.29\\
    HD 112357 & 304.4982 & 43.2255 & 75.3 &3.33\\
    HD 124683 & 329.4357 & 40.3105 & 80.2 & 3.34\\
    *87 Vir & 321.2814 & 43.0081 &193 &3.34\\
    *kap Vir &	333.5161 &47.703 & 70.6 & 3.35\\
    HD 120544 &	321.4239 & 40.8407 & 85.1	&3.36 \\
    HD 114523 & 310.1621 & 48.3512 & 197.5 & 3.36 \\
    HD 116061 &	312.6072 & 42.8171	& 83.4	& 3.37 \\
    *69 Vir &  315.4449 &46.0215 &79.3 &3.37\\
    HD 124627  & 345.435 & 58.3274	&187.5	&3.37 \\
    *alf Vir & 316.1123 &50.8446 &76.6 &3.38\\
    *63 Vir & 313.499 &44.4874	&89.9	&3.4 \\
    HD 121136 &	324.7477 & 45.5467 & 87.2	&3.4 \\
    *p Vir & 333.4651 &57.5692 &75.2 &3.4 \\
    HD 114482 & 309.6416 & 45.5256 & 93.8 &3.43 \\
    HD 124758 & 331.3303 &43.0609 & 139.5	&3.44 \\
    *1 Ser & 356.1271 & 49.2235	& 98.5	&3.45 \\
    HD 127888 &	354.5897 & 56.73 & 166.3	&3.45 \\
    HD 125922 & 349.4423 &57.8606 &159.8	&3.46 \\
    HD 121910 &	325.6982 & 43.9698 &134.7	&3.47 \\
    HD 112519 & 304.7744 & 40.0963 &126	&3.48 \\
    HD 112504 & 305.399	& 53.9328 &162	&3.48 \\
    *psi Vir & 304.135 &53.3259	 &160.5	&3.49 \\
    HD 118411 &318.4994	&45.0413 &147.1	&3.5 \\
    * 75 Vir &317.4714	&46.3361 &145.6	&3.51 \\
    HD 118826 &319.6339	&45.689	&143	&3.51 \\
    HD 110396 &	300.0701 &40.091 &166.6	&3.52 \\
    HD 115694 &311.7378	&42.3951 &149.9	&3.53 \\
    BD-18 3508  &305.0478 &43.6571	&151.6	&3.54 \\
    HD 124460 &347.2242	&59.8929 &117.9	&3.59 \\
    \hline
    \end{tabular}
    }

    \label{tab:stars-near-Spica}
\end{table}


\newcommand{\actaa}{Acta Astron.} 

\newcommand{\araa}{Annu. Rev. Astron. Astrophys.} 

\newcommand{\aar}{Astron. Astrophys. Rev.} 

\newcommand{\ab}{Astrobiol.} 

\newcommand{\aj}{Astron. J.} 

\newcommand{\apj}{Astrophys. J.} 

\newcommand{\apjl}{Astrophys. J. Lett.} 

\newcommand{\apjs}{Astrophys. J. Suppl. Ser.} 

\newcommand{\ao}{Appl. Opt.} 

\newcommand{\apss}{Astrophys. Space Sci.} 

\newcommand{\aap}{Astron. Astrophys.} 

\newcommand{\aapr}{Astron. Astrophys. Rev.} 

\newcommand{\aaps}{Astron. Astrophys. Suppl.} 

\newcommand{\baas}{Bull. Am. Astron. Soc.} 

\newcommand{\caa}{Chinese Astron. Astrophys.} 

\newcommand{\cjaa}{Chinese J. Astron. Astrophys.} 

\newcommand{\cqg}{Class. Quantum Gravity} 

\newcommand{\gal}{Galaxies} 

\newcommand{\gca}{Geochim. Cosmochim. Acta} 

\newcommand{\icarus}{Icarus} 

\newcommand{\jcap}{J. Cosmol. Astropart. Phys.} 

\newcommand{\jgr}{J. Geophys. Res.} 

\newcommand{\jgrp}{J. Geophys. Res.: Planets} 

\newcommand{\jqsrt}{J. Quant. Spectrosc. Radiat. Transf.} 

\newcommand{\memsai}{Mem. Soc. Astron. Italiana} 

\newcommand{\mnras}{Mon. Not. R. Astron. Soc.} 

\newcommand{\nat}{Nature} 

\newcommand{\nastro}{Nat. Astron.} 

\newcommand{\ncomms}{Nat. Commun.} 

\newcommand{\nphys}{Nat. Phys.} 

\newcommand{\na}{New Astron.} 

\newcommand{\nar}{New Astron. Rev.} 

\newcommand{\physrep}{Phys. Rep.} 

\newcommand{\pra}{Phys. Rev. A} 

\newcommand{\prb}{Phys. Rev. B} 

\newcommand{\prc}{Phys. Rev. C} 

\newcommand{\prd}{Phys. Rev. D} 

\newcommand{\pre}{Phys. Rev. E} 

\newcommand{\prl}{Phys. Rev. Lett.} 

\newcommand{\psj}{Planet. Sci. J.} 

\newcommand{\planss}{Planet. Space Sci.} 

\newcommand{\pnas}{Proc. Natl Acad. Sci. USA} 

\newcommand{\procspie}{Proc. SPIE} 

\newcommand{\pasa}{Publ. Astron. Soc. Aust.} 

\newcommand{\pasj}{Publ. Astron. Soc. Jpn} 

\newcommand{\pasp}{Publ. Astron. Soc. Pac.} 

\newcommand{\rmxaa}{Rev. Mexicana Astron. Astrofis.} 

\newcommand{\sci}{Science} 

\newcommand{\sciadv}{Sci. Adv.} 

\newcommand{\solphys}{Sol. Phys.} 

\newcommand{\sovast}{Soviet Ast.} 

\newcommand{\ssr}{Space Sci. Rev.} 

\newcommand{\uni}{Universe} 
\bibliographystyle{elsarticle-harv} 
\bibliography{references_1}

\begin{thebibliography}{30}
\expandafter\ifx\csname natexlab\endcsname\relax\def\natexlab#1{#1}\fi
\providecommand{\url}[1]{\texttt{#1}}
\providecommand{\href}[2]{#2}
\providecommand{\path}[1]{#1}
\providecommand{\DOIprefix}{doi:}
\providecommand{\ArXivprefix}{arXiv:}
\providecommand{\URLprefix}{URL: }
\providecommand{\Pubmedprefix}{pmid:}
\providecommand{\doi}[1]{\href{http://dx.doi.org/#1}{\path{#1}}}
\providecommand{\Pubmed}[1]{\href{pmid:#1}{\path{#1}}}
\providecommand{\bibinfo}[2]{#2}
\ifx\xfnm\relax \def\xfnm[#1]{\unskip,\space#1}\fi
\bibitem[{{Baes} et~al.(2022){Baes}, {Camps} and {Kapoor}}]{Baes2022}
\bibinfo{author}{{Baes}, M.}, \bibinfo{author}{{Camps}, P.}, \bibinfo{author}{{Kapoor}, A.U.}, \bibinfo{year}{2022}.
\newblock \bibinfo{title}{{A new analytical scattering phase function for interstellar dust}}.
\newblock \bibinfo{journal}{\aap} \bibinfo{volume}{659}, \bibinfo{pages}{A149}.
\newblock \DOIprefix\doi{10.1051/0004-6361/202142437}, \href{http://arxiv.org/abs/2202.01607}{{\tt arXiv:2202.01607}}.
\bibitem[{{Blum} and {Wurm}(2000)}]{Blum2000}
\bibinfo{author}{{Blum}, J.}, \bibinfo{author}{{Wurm}, G.}, \bibinfo{year}{2000}.
\newblock \bibinfo{title}{{Experiments on Sticking, Restructuring, and Fragmentation of Preplanetary Dust Aggregates}}.
\newblock \bibinfo{journal}{\icarus} \bibinfo{volume}{143}, \bibinfo{pages}{138--146}.
\newblock \DOIprefix\doi{10.1006/icar.1999.6234}.
\bibitem[{{Blum} et~al.(1998){Blum}, {Wurm}, {Poppe} and {Heim}}]{Blum1998}
\bibinfo{author}{{Blum}, J.}, \bibinfo{author}{{Wurm}, G.}, \bibinfo{author}{{Poppe}, T.}, \bibinfo{author}{{Heim}, L.O.}, \bibinfo{year}{1998}.
\newblock \bibinfo{title}{{Aspects of Laboratory Dust Aggregation with Relevance to the Formation of Planetesimals}}, in: \bibinfo{editor}{{Ehrenfreund}, P.}, \bibinfo{editor}{{Krafft}, C.}, \bibinfo{editor}{{Kochan}, H.}, \bibinfo{editor}{{Pirronello}, V.} (Eds.), \bibinfo{booktitle}{Laboratory astrophysics and space research}, p. \bibinfo{pages}{399}.
\newblock \DOIprefix\doi{10.1007/978-94-011-4728-6_15}.
\bibitem[{Carciofi et~al.(2007)Carciofi, Magalhães, Leister, Bjorkman and Levenhagen}]{Carciofi2007}
\bibinfo{author}{Carciofi, A.C.}, \bibinfo{author}{Magalhães, A.M.}, \bibinfo{author}{Leister, N.V.}, \bibinfo{author}{Bjorkman, J.E.}, \bibinfo{author}{Levenhagen, R.S.}, \bibinfo{year}{2007}.
\newblock \bibinfo{title}{Achernar: Rapid polarization variability as evidence of photospheric and circumstellar activity}.
\newblock \bibinfo{journal}{The Astrophysical Journal} \bibinfo{volume}{671}, \bibinfo{pages}{L49–L52}.
\newblock \URLprefix \url{http://dx.doi.org/10.1086/524772}, \DOIprefix\doi{10.1086/524772}.
\bibitem[{{Centurion} and {Vladilo}(1991)}]{Centurion1991}
\bibinfo{author}{{Centurion}, M.}, \bibinfo{author}{{Vladilo}, G.}, \bibinfo{year}{1991}.
\newblock \bibinfo{title}{{The Local Interstellar Medium toward the Center of Loop I}}.
\newblock \bibinfo{journal}{\apj} \bibinfo{volume}{372}, \bibinfo{pages}{494}.
\newblock \DOIprefix\doi{10.1086/169995}.
\bibitem[{Choi et~al.(2013)Choi, Min, Seon, Lim, Jo and Park}]{Choi2013}
\bibinfo{author}{Choi, Y.J.}, \bibinfo{author}{Min, K.W.}, \bibinfo{author}{Seon, K.I.}, \bibinfo{author}{Lim, T.H.}, \bibinfo{author}{Jo, Y.S.}, \bibinfo{author}{Park, J.W.}, \bibinfo{year}{2013}.
\newblock \bibinfo{title}{Far-ultraviolet observations of the spica nebula and the interaction zone}.
\newblock \bibinfo{journal}{The Astrophysical Journal} \bibinfo{volume}{774}, \bibinfo{pages}{34}.
\newblock \URLprefix \url{https://dx.doi.org/10.1088/0004-637X/774/1/34}, \DOIprefix\doi{10.1088/0004-637X/774/1/34}.
\bibitem[{{Demyk} et~al.(2022){Demyk}, {Gromov}, {Meny}, {Ysard}, {Paradis}, {Jones}, {Petitprez}, {Hubert}, {Leroux}, {Nayral} and {Delpech}}]{2022Demyk}
\bibinfo{author}{{Demyk}, K.}, \bibinfo{author}{{Gromov}, V.}, \bibinfo{author}{{Meny}, C.}, \bibinfo{author}{{Ysard}, N.}, \bibinfo{author}{{Paradis}, D.}, \bibinfo{author}{{Jones}, A.P.}, \bibinfo{author}{{Petitprez}, D.}, \bibinfo{author}{{Hubert}, P.}, \bibinfo{author}{{Leroux}, H.}, \bibinfo{author}{{Nayral}, C.}, \bibinfo{author}{{Delpech}, F.}, \bibinfo{year}{2022}.
\newblock \bibinfo{title}{{Low-temperature optical constants of amorphous silicate dust analogues}}.
\newblock \bibinfo{journal}{\aap} \bibinfo{volume}{666}, \bibinfo{pages}{A192}.
\newblock \DOIprefix\doi{10.1051/0004-6361/202243815}, \href{http://arxiv.org/abs/2209.06513}{{\tt arXiv:2209.06513}}.
\bibitem[{{Dorschner} et~al.(1995){Dorschner}, {Begemann}, {Henning}, {Jaeger} and {Mutschke}}]{Dorschner1995}
\bibinfo{author}{{Dorschner}, J.}, \bibinfo{author}{{Begemann}, B.}, \bibinfo{author}{{Henning}, T.}, \bibinfo{author}{{Jaeger}, C.}, \bibinfo{author}{{Mutschke}, H.}, \bibinfo{year}{1995}.
\newblock \bibinfo{title}{{Steps toward interstellar silicate mineralogy. II. Study of Mg-Fe-silicate glasses of variable composition.}}
\newblock \bibinfo{journal}{\aap} \bibinfo{volume}{300}, \bibinfo{pages}{503}.
\bibitem[{{Draine}(2003)}]{Draine2003_Scattering}
\bibinfo{author}{{Draine}, B.T.}, \bibinfo{year}{2003}.
\newblock \bibinfo{title}{{Scattering by Interstellar Dust Grains. I. Optical and Ultraviolet}}.
\newblock \bibinfo{journal}{\apj} \bibinfo{volume}{598}, \bibinfo{pages}{1017--1025}.
\newblock \DOIprefix\doi{10.1086/379118}, \href{http://arxiv.org/abs/astro-ph/0304060}{{\tt arXiv:astro-ph/0304060}}.
\bibitem[{Draine and Hensley(2021)}]{draine2021}
\bibinfo{author}{Draine, B.T.}, \bibinfo{author}{Hensley, B.S.}, \bibinfo{year}{2021}.
\newblock \bibinfo{title}{{The Dielectric Function of "Astrodust", and Absorption and Scattering Cross Sect ions for Spheroids}}.
\newblock \URLprefix \url{https://doi.org/10.34770/9ypp-dv78}.
\bibitem[{{Draine} and {Lee}(1984)}]{1984BTDraine}
\bibinfo{author}{{Draine}, B.T.}, \bibinfo{author}{{Lee}, H.M.}, \bibinfo{year}{1984}.
\newblock \bibinfo{title}{{Optical Properties of Interstellar Graphite and Silicate Grains}}.
\newblock \bibinfo{journal}{\apj} \bibinfo{volume}{285}, \bibinfo{pages}{89}.
\newblock \DOIprefix\doi{10.1086/162480}.
\bibitem[{{Fitzpatrick}(1999)}]{Fitzpatrick1999}
\bibinfo{author}{{Fitzpatrick}, E.L.}, \bibinfo{year}{1999}.
\newblock \bibinfo{title}{{Correcting for the Effects of Interstellar Extinction}}.
\newblock \bibinfo{journal}{\pasp} \bibinfo{volume}{111}, \bibinfo{pages}{63--75}.
\newblock \DOIprefix\doi{10.1086/316293}, \href{http://arxiv.org/abs/astro-ph/9809387}{{\tt arXiv:astro-ph/9809387}}.
\bibitem[{{Gontcharov} and {Mosenkov}(2019)}]{Gontcharov2019}
\bibinfo{author}{{Gontcharov}, G.A.}, \bibinfo{author}{{Mosenkov}, A.V.}, \bibinfo{year}{2019}.
\newblock \bibinfo{title}{{Interstellar polarization and extinction in the Local Bubble and the Gould Belt}}.
\newblock \bibinfo{journal}{\mnras} \bibinfo{volume}{483}, \bibinfo{pages}{299--314}.
\newblock \DOIprefix\doi{10.1093/mnras/sty2978}, \href{http://arxiv.org/abs/1811.01411}{{\tt arXiv:1811.01411}}.
\bibitem[{{Halder}(2022)}]{2022Halder}
\bibinfo{author}{{Halder}, P.}, \bibinfo{year}{2022}.
\newblock \bibinfo{title}{{REST: A Java Package for Crafting Realistic Cosmic Dust Particles}}.
\newblock \bibinfo{journal}{\apjs} \bibinfo{volume}{263}, \bibinfo{pages}{3}.
\newblock \DOIprefix\doi{10.3847/1538-4365/ac9183}, \href{http://arxiv.org/abs/2209.05768}{{\tt arXiv:2209.05768}}.
\bibitem[{{Hensley} and {Draine}(2023)}]{Hensley2023}
\bibinfo{author}{{Hensley}, B.S.}, \bibinfo{author}{{Draine}, B.T.}, \bibinfo{year}{2023}.
\newblock \bibinfo{title}{{The Astrodust+PAH Model: A Unified Description of the Extinction, Emission, and Polarization from Dust in the Diffuse Interstellar Medium}}.
\newblock \bibinfo{journal}{\apj} \bibinfo{volume}{948}, \bibinfo{pages}{55}.
\newblock \DOIprefix\doi{10.3847/1538-4357/acc4c2}, \href{http://arxiv.org/abs/2208.12365}{{\tt arXiv:2208.12365}}.
\bibitem[{{Henyey} and {Greenstein}(1941)}]{Henyey_Greenstein_1941}
\bibinfo{author}{{Henyey}, L.G.}, \bibinfo{author}{{Greenstein}, J.L.}, \bibinfo{year}{1941}.
\newblock \bibinfo{title}{{Diffuse radiation in the Galaxy.}}
\newblock \bibinfo{journal}{\apj} \bibinfo{volume}{93}, \bibinfo{pages}{70--83}.
\newblock \DOIprefix\doi{10.1086/144246}.
\bibitem[{{Jones} et~al.(2017){Jones}, {K{\"o}hler}, {Ysard}, {Bocchio} and {Verstraete}}]{2017Jones}
\bibinfo{author}{{Jones}, A.P.}, \bibinfo{author}{{K{\"o}hler}, M.}, \bibinfo{author}{{Ysard}, N.}, \bibinfo{author}{{Bocchio}, M.}, \bibinfo{author}{{Verstraete}, L.}, \bibinfo{year}{2017}.
\newblock \bibinfo{title}{{The global dust modelling framework THEMIS}}.
\newblock \bibinfo{journal}{\aap} \bibinfo{volume}{602}, \bibinfo{pages}{A46}.
\newblock \DOIprefix\doi{10.1051/0004-6361/201630225}, \href{http://arxiv.org/abs/1703.00775}{{\tt arXiv:1703.00775}}.
\bibitem[{{Jones} et~al.(2011){Jones}, {West} and {Foster}}]{Jones2011_White_Dwarf_Spectra}
\bibinfo{author}{{Jones}, D.O.}, \bibinfo{author}{{West}, A.A.}, \bibinfo{author}{{Foster}, J.B.}, \bibinfo{year}{2011}.
\newblock \bibinfo{title}{{Using M Dwarf Spectra to Map Extinction in the Local Galaxy}}.
\newblock \bibinfo{journal}{\aj} \bibinfo{volume}{142}, \bibinfo{pages}{44}.
\newblock \DOIprefix\doi{10.1088/0004-6256/142/2/44}, \href{http://arxiv.org/abs/1102.0280}{{\tt arXiv:1102.0280}}.
\bibitem[{Kervella et~al.(2008)Kervella, Domiciano~de Souza, Kanaan, Meilland, Spang and Stee}]{Kervella2008}
\bibinfo{author}{Kervella, P.}, \bibinfo{author}{Domiciano~de Souza, A.}, \bibinfo{author}{Kanaan, S.}, \bibinfo{author}{Meilland, A.}, \bibinfo{author}{Spang, A.}, \bibinfo{author}{Stee, P.}, \bibinfo{year}{2008}.
\newblock \bibinfo{title}{The environment of the fast rotating star achernar: Ii. thermal infrared interferometry with vlti/midi}.
\newblock \bibinfo{journal}{Astronomy \&; Astrophysics} \bibinfo{volume}{493}, \bibinfo{pages}{L53–L56}.
\newblock \URLprefix \url{http://dx.doi.org/10.1051/0004-6361:200810980}, \DOIprefix\doi{10.1051/0004-6361:200810980}.
\bibitem[{{Kurucz}(1979)}]{Kurucz1979_model_atmospheres}
\bibinfo{author}{{Kurucz}, R.L.}, \bibinfo{year}{1979}.
\newblock \bibinfo{title}{{Model atmospheres for G, F, A, B, and O stars.}}
\newblock \bibinfo{journal}{ApJs} \bibinfo{volume}{40}, \bibinfo{pages}{1--340}.
\newblock \DOIprefix\doi{10.1086/190589}.
\bibitem[{{Mathis} and {Whiffen}(1989)}]{Mathis1989}
\bibinfo{author}{{Mathis}, J.S.}, \bibinfo{author}{{Whiffen}, G.}, \bibinfo{year}{1989}.
\newblock \bibinfo{title}{{Composite Interstellar Grains}}.
\newblock \bibinfo{journal}{\apj} \bibinfo{volume}{341}, \bibinfo{pages}{808}.
\newblock \DOIprefix\doi{10.1086/167538}.
\bibitem[{{Murthy} and {Henry}(2011)}]{Murthy2011_UV_halos}
\bibinfo{author}{{Murthy}, J.}, \bibinfo{author}{{Henry}, R.}, \bibinfo{year}{2011}.
\newblock \bibinfo{title}{{Dust-scattered Ultraviolet Halos around Bright Stars}}.
\newblock \bibinfo{journal}{\apj} \bibinfo{volume}{734}, \bibinfo{pages}{13}.
\newblock \DOIprefix\doi{10.1088/0004-637X/734/1/13}, \href{http://arxiv.org/abs/1011.5982}{{\tt arXiv:1011.5982}}.
\bibitem[{{Park} et~al.(2010){Park}, {Min}, {Seon}, {Han} and {Edelstein}}]{Park2010_HII_region}
\bibinfo{author}{{Park}, J.W.}, \bibinfo{author}{{Min}, K.W.}, \bibinfo{author}{{Seon}, K.I.}, \bibinfo{author}{{Han}, W.}, \bibinfo{author}{{Edelstein}, J.}, \bibinfo{year}{2010}.
\newblock \bibinfo{title}{{Analysis of Spatial Structure of the Spica H II Region}}.
\newblock \bibinfo{journal}{\apj} \bibinfo{volume}{719}, \bibinfo{pages}{1964--1968}.
\newblock \DOIprefix\doi{10.1088/0004-637X/719/2/1964}, \href{http://arxiv.org/abs/1007.0802}{{\tt arXiv:1007.0802}}.
\bibitem[{{Puspitarini} and {Lallement}(2012)}]{Puspitarini2012}
\bibinfo{author}{{Puspitarini}, L.}, \bibinfo{author}{{Lallement}, R.}, \bibinfo{year}{2012}.
\newblock \bibinfo{title}{{Distance to northern high-latitude HI shells}}.
\newblock \bibinfo{journal}{\aap} \bibinfo{volume}{545}, \bibinfo{pages}{A21}.
\newblock \DOIprefix\doi{10.1051/0004-6361/201219284}, \href{http://arxiv.org/abs/1207.5353}{{\tt arXiv:1207.5353}}.
\bibitem[{{Shalima} et~al.(2013){Shalima}, {Murthy} and {Gupta}}]{Shalima2013}
\bibinfo{author}{{Shalima}, P.}, \bibinfo{author}{{Murthy}, J.}, \bibinfo{author}{{Gupta}, R.}, \bibinfo{year}{2013}.
\newblock \bibinfo{title}{{Dust properties from GALEX observations of a UV halo around Spica}}.
\newblock \bibinfo{journal}{Earth, Planets and Space} \bibinfo{volume}{65}, \bibinfo{pages}{1123--1126}.
\newblock \DOIprefix\doi{10.5047/eps.2013.05.020}, \href{http://arxiv.org/abs/1307.6832}{{\tt arXiv:1307.6832}}.
\bibitem[{{Shen} et~al.(2008){Shen}, {Draine} and {Johnson}}]{2008YueShen}
\bibinfo{author}{{Shen}, Y.}, \bibinfo{author}{{Draine}, B.T.}, \bibinfo{author}{{Johnson}, E.T.}, \bibinfo{year}{2008}.
\newblock \bibinfo{title}{{Modeling Porous Dust Grains with Ballistic Aggregates. I. Geometry and Optical Properties}}.
\newblock \bibinfo{journal}{\apj} \bibinfo{volume}{689}, \bibinfo{pages}{260--275}.
\newblock \DOIprefix\doi{10.1086/592765}, \href{http://arxiv.org/abs/0801.1996}{{\tt arXiv:0801.1996}}.
\bibitem[{{Witt} and {Petersohn}(1994)}]{Witt1994}
\bibinfo{author}{{Witt}, A.N.}, \bibinfo{author}{{Petersohn}, J.K.}, \bibinfo{year}{1994}.
\newblock \bibinfo{title}{{Scattering by Diffuse Clouds in the Galaxy: Searching for the Cosmic Background}}, in: \bibinfo{editor}{{Cutri}, R.M.}, \bibinfo{editor}{{Latter}, W.B.} (Eds.), \bibinfo{booktitle}{The First Symposium on the Infrared Cirrus and Diffuse Interstellar Clouds}, p.~\bibinfo{pages}{91}.
\bibitem[{{Zagury} et~al.(1998){Zagury}, {Jones} and {Boulanger}}]{Zagury1998_Dust_composition}
\bibinfo{author}{{Zagury}, F.}, \bibinfo{author}{{Jones}, A.}, \bibinfo{author}{{Boulanger}, F.}, \bibinfo{year}{1998}.
\newblock \bibinfo{title}{Dust composition in the low density medium around spica}.
\newblock \bibinfo{journal}{International Astronomical Union Colloquium} \bibinfo{volume}{506}, \bibinfo{pages}{385--388}.
\newblock \DOIprefix\doi{10.1007/BFb0104751}.
\bibitem[{Zaninetti(2020)}]{Zaninetti2020_local_bubble_shape}
\bibinfo{author}{Zaninetti, L.}, \bibinfo{year}{2020}.
\newblock \bibinfo{title}{On the shape of the local bubble}.
\newblock \bibinfo{journal}{International Journal of Astronomy and Astrophysics} \bibinfo{volume}{10}, \bibinfo{pages}{11--27}.
\newblock \DOIprefix\doi{10.4236/ijaa.2020.101002}.
\bibitem[{{Zubko} et~al.(2006){Zubko}, {Shkuratov}, {Kiselev} and {Videen}}]{2006Zubko}
\bibinfo{author}{{Zubko}, E.}, \bibinfo{author}{{Shkuratov}, Y.}, \bibinfo{author}{{Kiselev}, N.N.}, \bibinfo{author}{{Videen}, G.}, \bibinfo{year}{2006}.
\newblock \bibinfo{title}{{DDA simulations of light scattering by small irregular particles with various structure}}.
\newblock \bibinfo{journal}{\jqsrt} \bibinfo{volume}{101}, \bibinfo{pages}{416--434}.
\newblock \DOIprefix\doi{10.1016/j.jqsrt.2006.02.055}.

\end{thebibliography}







\end{document}